\documentclass[doublespacing]{elsart}

\begin{document}
\runauthor{Cicero, Caesar and Vergil}
\begin{frontmatter}
\title{Optical Alignment System for the PHENIX Muon Tracking Chambers}
\author[RIKEN,RBRC]{J. Murata\thanksref{jiro}}
\author[NMSU]{A. Al-Jamel}
\author[NMSU]{R.L. Armendariz}
\author[LANL]{M.L. Brooks}
\author[TITECH]{T. Horaguchi}
\author[TITECH]{N. Kamihara}
\author[RBRC]{H. Kobayashi}
\author[LANL]{D.M. Lee}
\author[TITECH]{T.-A. Shibata}
\author[LANL]{W.E. Sondheim}
\thanks[jiro]{e-mail: jiro@riken.jp}

\address[RIKEN]{RIKEN, Wako 351-0198, Japan}
\address[RBRC]{RIKEN BNL Research Center, Brookhaven National
 Lab., Upton, NY 11973}
\address[NMSU]{New Mexico State University, Las Cruces, NM 88003}
\address[LANL]{Los Alamos National Lab. P.O. Box 1663, Los Alamos, NM 87545}
\address[TITECH]{Department of Physics, Tokyo Institute of Technology, Tokyo 152-8551, Japan}
\begin{abstract}
A micron-precision optical alignment system (OASys) for the PHENIX muon 
tracking chambers is developed. To ensure the required mass resolution of
 vector meson detection, the relative alignment between three tracking station
 chambers must be monitored with a precision of 25$\mu$m. 
The OASys is a straightness monitoring system comprised of a light
 source, lens and CCD camera, 
used for determining the initial placement
as well as for 
monitoring the time dependent movement of the chambers on a micron scale. 
\end{abstract}
\end{frontmatter}
\section{Introduction}
The PHENIX experiment at RHIC has two muon arms (north and south arms)
placed at the forward rapidity regions.
The role of the PHENIX muon arms is to track and identify muons,
providing good rejection of pions and kaons.
To accomplish this, we employ a radial field magnetic
spectrometer with precision muon tracking chambers followed by a stack
of absorber/low-resolution tracking layers (muon identifier), as 
shown in Fig.\ref{muonarm}. 
The muon tracking chambers consist of three sets of cathode
strip readout tracking chambers spanning 4.5(3.5)m for the north(south)
arm, mounted inside conical magnets, with multiple cathode orientations
and readout planes in each station \cite{PHENIX-NIM,INPC}.
The main goals of the muon arms are to study vector meson production,
the Drell-Yan process, heavy quark production, and $Z^0$, $W^{\pm}$
production at the forward rapidities.
Clean separation of $J/\Psi$ from $\Psi'$ requires a muon-detection position
resolution of 60$\mu$m at each station.
To maintain the momentum resolution, an optical alignment system is
installed to calibrate the initial placement of the chambers, and to monitor 
thermal expansion of the chambers. 
\begin{figure}[p]
\begin{center}
\caption{One arm of the PHENIX muon detector. Three stations of the
tracking chambers are placed inside the muon magnet. Each station
 consists of eight octant cathode strip chambers.}
\end{center}
\label{muonarm}
\end{figure}
\section{Optical Alignment System (OASys)}
The muon momentum is determined by measuring the displacement of a muon
hit position at the station two chamber with respect to a straight line between those at stations one and three; therefore, only the relative straightness must be known to high accuracy.
The absolute placement of the chambers is surveyed with respect to a
PHENIX hall monument system and has accuracy of 1-2mm.
The absolute positions of the chambers need to be known only to a few
mm, but the relative alignment of the chambers to each other
must be known to much higher precision than the chamber resolution.
One prime task of the OASys is to measure the relative alignment after installing the chambers in the muon magnet.
In addition, real-time monitoring of the relative straightness is required for correcting the motions of the chambers.
To fulfill this requirement, the OASys must be able to measure the
chamber position in the time span of minutes anytime during data taking
within a 25$\mu$m accuracy.
The requirement on the OASys accuracy is obtained by a simulation studies described in \cite{IEEE}.

The OASys is a straightness monitor consisting of a fiber optic
divergent light source at station one, a convex lens at station two, and a CCD camera at station three.
Any deflection at station 1, 2, or 3 causes a displacement of the focal center from the center point on the CCD with a magnification factor of around 2.5 for our system.
If the measured deflection is always attributed to the movement of station 2, then the relative misalignment of the three stations is accounted for.
There are seven OASys beams surrounding each octant chamber, therefore,
there are $7\times8=56$ OASys beams in total for one muon arm, as shown in Fig.\ref{OASYS}.
By combining the seven OASys beams for one octant chamber, the translational misalignment of the chambers, as well as rotational and linear temperature expansions, can be measured.
Similar systems were used in L3 \cite{L3}, proposed for the GEM SSC
Muon System \cite{GEM}, and designed for the ATLAS muon spectrometer
\cite{ATLAS}.
The present system measures the relative chamber positions only in X-Y
directions (within the plane  parallel to the chamber plane).
The contribution of the misalignment in the Z direction to the momentum 
resolution is negligible to that of
misalignment in X-Y directions.
Therefore, an initial placement with an accuracy of about 2 mm is sufficient for our system.

The light source block consists of a fiber optic cable, cable terminator and mounting block.
The mounting block is precisely pinned to the chamber frame and the
cable terminator is located in the mounting block by a special procedure
described in Sec.\ref{alignment}.
We use a 15m-long, core/cladding diameter = 62.5/125$\mu$m multi-mode
fiber optic cable with an FC connector at the chamber-side end, and bare fiber finish at the light source box side.
The custom-made mounting block is designed to hold and tune the position of the FC fiber connector.
A bundle of 56 fiber optic cables is connected to the light source box
(FiberPro150, High Sierra Lighting).
A 150 watt metal halide high-intensity discharge lamp with an average life of 6000 hours is used.

The lens block consists of a convex lens, a lens holder with an XY translation stage, and a mounting block.
Because the OASys beams surround the chamber frame, they have different focal lengths.
To accommodate the various focal lengths, 
we have chosen to use single
commercial lenses with focal lengths differing in steps of 100mm.
A typical distance between the light source and lens is about
$l_1=1200$mm, and between the lens and CCD camera is about $l_2=700$mm.
The required focal length $f$ of the lens is determined by $1/f=1/l_1+1/l_2$.
A typical focal length for our system is about 700mm.
We use 1cm diameter plano convex glass lenses (MELLES GRIOT, Plano-Convex Glass Lens) of 600, 700, and 800mm focal lengths.
The use of this series of lenses enables most of the OASys beams to
produce a single sufficiently sharp focal image at the CCD position.
However, some of the OASys beams with poorly focused lenses produce a concentric ring interference pattern on the CCD image.
Some of them even exhibit destructive interference at the center point.
These poorly focused OASys beams have broad, relatively low-intensity focal
images.
However, the center positions of the broad images are determined
well by our readout DAQ system described in Sec.\ref{Sec-DAQ}.
The lenses are mounted inside a small ring-like lens-holding cell (MELLES GRIOT, Optical Component Cell).
The placement of the lens into 
the lens holding cell is achieved by using 
an XY positioning device to center the lens.
Precision pins accurately hold the lens holding cell in the lens block
and the lens block is accurately pinned to the chamber frame.
The lens-holding cells are set inside the XY positioning block (MELLES
GRIOT, Y-Z Positioner for Optical Component Cells) which is attached on the mounting block with precision alignment pins.
The same as for the light source block, the mounting block is placed on the chamber frame using precision alignment pins.

The CCD camera block consists of the CCD camera, camera holder and mounting block.
The use of precision alignment pins ensures connection between the camera holder and the mounting block, and between the mounting block and the station three chamber frame.
Considering the possible initial misplacement of the chamber position
inside the muon magnet, the expected center position of the focal image
on the CCD camera can be displaced from the camera center point by
distances of a few mm.
To allow for the possibility of a wide dynamic position range,
we use a CCD camera (HITACHI DENSHI, KP-M1U) that has
a $8.8\times 6.6$mm (768(H)$\times$493(V) pixels) effective region.
The pixel size is 11.0(H)$\mu$m$\times$13.0(V)$\mu$m.
The video signal is sent through the EIA video format to the DAQ system.
\begin{figure}[p]
\begin{center}
\caption{Overview of the OASys. Seven OASys beams, from the light
 source, through the lens to the CCD camera, surround one octant chamber.
There are $7\times8=56$ OASys beams in total for one muon arm. 
Relative straightness is measured from the focal image position on the CCD camera.}
\end{center}
\label{OASYS}
\end{figure}
\section{System Alignment}
\label{alignment}
Each component of the OASys must be placed precisely at the designed
position. 
The required accuracy of placement is better than 25$\mu$m. 
For this purpose, we built an OASys-component-alignment system with a small optical rotational stage. 
We set a fiber optic light source, a convex lens and a CCD camera at a well-focused position on a 4m-long optical table.
The light-source-mounting block with the fiber optic cable is placed on
the precision optical rotational stage, which has the same alignment-pin holes as that on the chamber frame.
In order to tune the relative fiber optic cable position to the mounting
block, the mounting block is rotated around the optic beam axis using
the rotational table, while monitoring the movement of the focal image on the fixed CCD camera.
A peak position movement within about one CCD pixel is required in order to satisfy the positioning requirement.
Lens position alignment on the lens holders is performed the same as for
light source alignment; the lens holders are rotated using the
rotational table and fixed light source and CCD camera.
In the case of camera position alignment, we measure the relative position of the camera with respect to the camera-mounting block instead of tuning the camera position.
The camera position is measured for each CCD camera by rotating the
camera-mounting block, while monitoring the peak position of the focal image.
The focal positions measured while rotating the CCD camera form a ring-image trace scatter plot.
The center point of the camera-mounting block is determined to 10$\mu$m by circle fitting on the ring-like
distribution.

Final positions of the OASys components after the installation onto the
chambers can be calibrated by an off-line analysis comparing all the
OASys data, and by a real experimental data analysis for the straight line 
tracks obtained without applying the magnetic field, as well as for 
the geometry alignments of the chambers.
\section{Data Acquisition and Online Data Analysis}
\label{Sec-DAQ}
The block diagram of the CCD readout DAQ system is shown in Fig.\ref{DAQ}.
The 56 CCD camera signals are fed into a video-signal switching
multiplexer (Keithley 7001+7011S) which has one video-signal output
port. The multiplexer Keithley 7001 has two slots, in which we use two 40-channel multiplexer cards (Keithley 7011-S).
CCD channel switching is controlled by external TTL signals.
The TTL channel switching signal cable and the video-signal cable are connected between the multiplexer and the frame grabber card on the DAQ PC.
In order to separate the ground level of the multiplexer located near the chambers and that of the DAQ PC, both signals are transmitted via optical fibers.
\begin{figure}[b]
\begin{center}
\caption{Block diagram of the CCD readout DAQ system. 
CCD channel-switching system is controlled by the same system.}
\label{DAQ}
\end{center}
\end{figure}
A simple video signal transmitter/receiver (Black Box, FiberPath AC444A)
is used for the video signal as well as for the TTL control signal for this purpose.
We use a frame grabber card (Scion Corporation, LG-3) for the PCI bus on a DAQ PC running on Windows 2000.
LG-3 has four monochrome video inputs with 640$\times$480 pixels, and four TTL input and output ports.
Only inside region of the 640$\times$480 pixels of
the EIA video signals with 768(H)$\times$493(V) pixels from the CCD
cameras can be captured using LG-3.
We use a single video input channel, sorting all 56 channels by sending
TTL switching signals from the digital output ports to the multiplexer.
All 56 camera images are captured every hour and their peak positions
are recorded automatically by the auto-semi-online analysis system.
For image processing, we use image analyzing software (Sion Image ver.$\beta$4.02 for Windows, Sion Corporation) which can also directly handle hardware devices of the LG-3.
Single PASCAL-like macro-script running on the Sion Image executes the following.
\begin{enumerate}
\item Select CCD channel by sending TTL control signals
\item Capture video images and store them into two-dimensional data array
\item Search for the highest intensity pixel point
\item Roughly determine peak position by calculating center of gravity position around the highest intensity pixel point
\item Generate one-dimensional light intensity histograms for X and Y
      directions, by projecting the intensity image with a narrow cut around the center of gravity
\end{enumerate}
These image analyzing procedures including hardware controls are very
stable because they are free of any I/O conflicting problems which may occur when we use combined systems. 
One OASys data-taking cycle, including 56 CCD channel scanning, is
executed every hour in order to monitor fine chamber movement,
thermal expansion and magnetic distortion.
After scanning all the channels, the stored 56 X-directional histograms and 56 Y-directional histograms are sent to a Linux PC via Ethernet as ASCII files.

On the Linux PC, using the ROOT framework \cite {root}, we perform fine peak-position determination with the histograms.
Subsequently semi-online analysis is executed automatically by a single ROOT macro-script.
\begin{enumerate}
\item Simple Gaussian fitting without any cut on the projected histograms
\item Determine ``window cut'' (center $\pm 0.7\sigma$ ) around the Gaussian peak region
\item Final Gaussian fitting ignoring the inside region of the window cut
\end{enumerate}
The reason why we must ignore the center region is that, sometimes particularly for those channels which have broad focal images, the peak shapes do not have simple Gaussian-like sharp peaks, while outer tails are well represented by Gaussian shapes.
This is because a multi-ring interference pattern appears at a defocused position.
Although the peak spot size is sometimes more than a few mm wide, we can determine the peak position by the ``window'' fitting procedure with micron precision.
The final peak positions, their history plots, sliced histograms and the CCD images are monitored on an online WEB page.
\section{Results}
Sample CCD images are shown in Fig.\ref{sharp} and Fig.\ref{broad}. 
Fig.\ref{sharp} represents CCD images for well focused OASys beam
channels with sharp focal images. 
In contrast, Fig.\ref{broad} represents that for weakly focused channels with typical broad focal images.
Because the focal image intensity for the broad images is very low, it is difficult to recognize even the existence of the focal image.
In Fig.\ref{broad}, we can see a clear multi-ring interference pattern at the intensified figure.
\begin{figure}[p]
\begin{center}
\caption{Sample of focal image on the CCD, for a typical sharp-focus
 channel.
Sliced X- and Y-dimensional light intensity histograms are also shown beside the image. The solid curves are the results of Gaussian fitting.
The ``window cut'' regions ignored by the Gaussian fittings are indicated as the gray regions around the peaks.}
\label{sharp}
\end{center}
\end{figure}
\begin{figure}[p]
\begin{center}
\caption{Sample of focal image on the CCD for a typical broad-focus channel. 
Because the raw CCD image is too weak for the focal image to be
 recognized, an intensified image is also shown at the bottom left corner. 
The sliced X- and Y-dimensional light intensity histograms show that the wide, weak peaks can be clearly identified.
The window cut indicated by the gray region is indispensable because of its highly deformed peak shape.}
\label{broad}
\end{center}
\end{figure}
The X-dimensional and Y-dimensional sliced histograms generated by the DAQ PC are also shown beside the image.
Final Gaussian fitting curves and the window cut regions are also drawn.
In spite of the small S/N ratio for the broad focus image, the wide and
low peak is well identified by our system.
Since these systems are located inside the muon magnet surrounded by
steel plates with small cable throughput holes, there is still room to further reduce background light by shielding the cable holes.
The image intensity can also be improved by modifying the light source box and the fiber distributor.
However, our results show that the current system is satisfactory. 

We measured the focal position resolution as shown in Fig.\ref{resolution}. 
Peak position distributions for 1000 samples taken within 30 minutes for
the typical sharp channel and for the typical broad channel are displayed.
The measured resolution is 1.4$\mu$m for the sharp channel, and 3.1$\mu$m for the broad channel.
Considering the required focal position resolution of 25$\mu$m, the results are excellent.
\begin{figure}[p]
\begin{center}
\caption{Peak position distributions for the focal position resolution
 measurements: a) typical sharp-focus channel and b) typical broad-focus
 channel. One thousand events are taken for each channel.
The width $\sigma$ of the Gaussian-like peaks indicate the position determination resolution of 1.4 and 3.1$\mu$m.}
\label{resolution}
\end{center}
\end{figure}
The system has been running for almost one year in a stable
operation since its installation for the south muon arm early in 2000,
taking data every hour.
Fig.\ref{history} shows an example of the history plot of the peak
position over a period of 18 days.
Both horizontal (azimuthal) and vertical (radial) movements are plotted in Fig.\ref{history}.
They are well correlated with each other, indicating that deformation is parallel to a direction between horizontal and vertical axes at the OASys beam position.
The magnitudes of the deformations are about 10$\mu$m for the horizontal
direction and 50$\mu$m for the vertical direction, which represents large radial deformation.
By combining the peak position data from all the OASys beam channels, 
we can analyze the chamber deformation mode.
Such off-line analysis aimed at improvements of the muon-detecting
position resolution within the required precision is on-going.
Most of the focal position movements can be understood to be a result of
temperature change.
In Fig.\ref{history}, the room temperature measured inside the muon magnet is also plotted.
A clear correlation between the temperature and the peak positions can be found.
Although an air-conditioning system is used inside the experimental
hall, the room temperature varies from about 21$^\circ$C to
22.5$^\circ$C.
The results show that the temperature dependence of the movements is about 70$\mu$m/$^\circ$C.
Considering the possible large change of air temperature without air
conditioning, it can be said that temperature control is a key for
maintaining precise chamber geometries.
It should be mentioned that the peak position also depends on the 
magnetic field inside the muon magnet.
It is mainly because the temperature depends on the magnetic field
condition.
Therefore, stable magnet condition is also required.

The OASys system will also be used for the north muon arm which 
is scheduled for completion in 2002.
We modified the lens part for the north muon arm.
Instead of using the commercial lens with focal lengths differing in
steps of 100mm, we use custom made lenses with exact required focal
lengths.
By this improvement, all the OASys beams have sharp focal image
providing better position resolution.  
\begin{figure}[p]
\begin{center}
\caption{Time dependent motions of the relative focal positions are shown for a
 period of 18 days. 
Both horizontal and vertical directional focal positions are plotted
 with arbitrary position offsets. 
Room temperatures measured near the chambers are also plotted.}
\label{history}
\end{center}
\end{figure}
\section{Summary}
The development of a micron-precision optical alignment system (OASys)
consisting of a light source, lens, and CCD camera for monitoring straightness is described.
The OASys is built and installed within a precision of 25$\mu$m for the
muon tracking chambers of the south muon arm used in the PHENIX
experiment at RHIC in 2001-2002. Thermal movements of the chambers have been successfully monitored every hour for one year with 1-3$\mu$m precision.
As well as for the south muon arm, the north muon arm, with an improved
OASys, is going to start taking data soon.

\vspace{12pt}
{\bf Acknowledgements}\\
The authors appreciate R. Savino and BNL technical staffs for the
technical support on the development of the OASys mechanics. 
We also thank E.G. Romero from Hytec Inc. and D. Clark from Los Alamos
National Laboratory for their efforts on designing the mounting devices. 
Some of the authors acknowledge the support of RIKEN Special Postdoctoral
Researchers Program and Technology Research Associate Program.


\begin{thebibliography}{999}
\bibitem{PHENIX-NIM} H. Akikawa et al., ``PHENIX Muon Arms'', to be
	published in Nucl. Instr. Meth. {\bf A}
\bibitem{INPC} J. Murata et al., AIP Conf. Proc. {\bf Vol. 610} (2002) 947.
\bibitem{IEEE} M.L. Brooks, D.M. Lee, and W. Sondheim, {\em IEEE
	Trans. Nucl. Sci.} {\bf Vol. 44 No.3} (1997) 683.
\bibitem{L3} B. Adeva et al., Nucl. Instr. Meth. {\bf A289} (1990) 335.
\bibitem{GEM} G. Mitselmakher and A. Ostapchuk, SSC Report GEM TN-92-202.
\bibitem{ATLAS} A. Airapetian et al., ATLAS Muon Spectrometer Technical
	Design Report, CERN/LHCC/97-22, May 1997.
\bibitem{root} see http://root.cern.ch
\end{thebibliography}
\end{document}